\documentclass[10pt, conference, letterpaper]{IEEEtran}
\usepackage{cite}

\ifCLASSINFOpdf
\else
\fi
\usepackage{graphicx}
\usepackage{subfig}

\usepackage{amsfonts} 
\usepackage{upgreek} 
\usepackage[cmex10]{amsmath}
\usepackage{algorithm}
\usepackage{algorithmicx}
\usepackage{algpseudocode}

\usepackage{array}

\usepackage{mathrsfs}
\usepackage{bm}
\usepackage{subeqnarray}
\usepackage{cases}

\usepackage{booktabs}
\usepackage{multirow}

\usepackage{amsmath,amsthm,amssymb,amsfonts}
\makeatletter
\thm@headfont{\sc}
\makeatother

\begin{document}
%
\title{Robust Beamforming for Downlink 3D-MIMO Systems with $l_1$-norm Bounded CSI Uncertainty}

\author{
\IEEEauthorblockN{Kai Liu\IEEEauthorrefmark{1},
Hui Feng\IEEEauthorrefmark{1}, Tao Yang\IEEEauthorrefmark{1} and Bo Hu\IEEEauthorrefmark{1}\IEEEauthorrefmark{2}}
\IEEEauthorblockA{
\IEEEauthorrefmark{1}Department of Electronic Engineering\\
\IEEEauthorrefmark{2}Key Laboratory of EMW Information (MoE)\\
Fudan University, Shanghai, 200433, China\\
Emails: \{10110720016, hfeng, taoyang, bohu\}@fudan.edu.cn}
}
\maketitle

\begin{abstract}
In this paper, a novel robust beamforming scheme is proposed in three dimensional multi-input multi-output (3D-MIMO) systems. As one of the typical deployments of massive MIMO, a 3D-MIMO system owns sparse channels in angular domain. Thus, various of sparse channel estimation algorithms produce sparse channel estimation errors which can be utilized to narrow down the perturbation region of imperfect CSI. We investigate a $l_1$-norm bounded channel uncertainty model for the robust beamforming problems, which captures the sparse nature of channel errors. Compared with the conventional spherical uncertainty, we prove that the scheme with $l_1$-norm bounded uncertainty consumes less beamforming power with the same signal to interference and noise ratio (SINR) thresholds. The proposed scheme is reformulated as a second-order cone programming (SOCP) and simulation results verify the effectiveness of our algorithm.
\end{abstract}


%
\IEEEpeerreviewmaketitle

\section{Introduction}

Recently, 3D-MIMO has attracted significant attention in wireless communication systems \cite{fdmimo}. As one of the candidate implementations of massive MIMO \cite{ncwun}, the beamforming of 3D-MIMO can be designed in full 3D space, which can substantially improve the system capacity and alleviate the multi-user interference \cite{com3d}. The performance of beamforming relies heavily on precise CSI. However, in realistic scenarios, the channel knowledge is generally imperfect. Therefore, CSI uncertainty should be considered in beamforming design so that the system performance is robust to imperfect channels.

There are various of CSI uncertainty models in literature. The most commonly used CSI uncertainty sets are spherical or ellipsoidal region \cite{sph1,sph3,elli1quan1}, defined by $l_2$-norm or Frobenius norm. Another kind of CSI uncertainty set is a hyper-rectangular region representing quantization errors \cite{sph2quan2}, defined by infinite norm for real and imaginary part of CSI errors independently. A unified representation of the aforementioned CSI uncertainty models is proposed in \cite{inteelli2}, which coverers CSI uncertainty by intersection of multiple ellipsoids. In \cite{severalunc}, a unified framework is proposed to cover most commonly used uncertainty models.

However, the aforementioned CSI uncertainty models are not suitable for 3D-MIMO systems owning to the sparseness of 3D-MIMO channels. In physical propagation environment, the number of local scatterers is limited and the scale of 3D-MIMO channel is large. Thus, 3D-MIMO channels exhibit sparse property when transformed to angular domain \cite{ccsds}, for which various of sparse channel estimation schemes have been proposed to reduce the pilot overhead, complexity, and estimation errors \cite{sce1,ccsds,ssce}. By subtracting the estimated sparse channel from the true channel, the CSI estimation errors tend to be sparse as well. Thus, the conventional spherical uncertainty model cannot describe the channel errors accurately.

In this letter, we investigate a novel robust beamforming scheme in 3D-MIMO systems with $l_1$-norm bounded uncertainty. The goal is to minimize the beamforming power such that the worst case SINR targets are satisfied. The main contribution of this paper is to introduce a new uncertainty model representing sparse channel errors, which has rarely been considered for robust beamforming before. Since the $l_1$-norm bounded uncertainty region is smaller than the conventional spherical uncertainty region for a given uncertainty bound, the proposed scheme is proved to outperforms the scheme with spherical bounded uncertainty. The robust beamforming problem can be reformulated as an SOCP, which can be solved by convex optimization toolbox efficiently. Simulations results show that the proposed scheme can achieve the worst case SINR targets with less power than convention robust beamforming method.

\emph{Notations}: In this letter, we use lower case for scalars, lower case with bold font for vectors, upper case with bold font for matrices. Superscripts ${\left( \cdot \right)^*}$, and ${\left( \cdot \right)^H}$ denote the complex conjugate and Hermitian transpose respectively. ${\left\| {\bf{a}} \right\|}_p$ denotes the $l_p$ norm of vector $\bf{a}$. $vec(\cdot)$ denotes stacking all columns of a matrix in a vector. ${{\mathbb{C}}^{m \times n}}$ is the set of ${m \times n}$ matrices in complex field.

\section{Structured Sparseness of 3D-MIMO Channels}

Consider a single-cell downlink multi-user 3D-MIMO system. The base station (BS) is equipped with a uniform planar array (UPA) consisting of $N_{\rm{v}}$ antennas in the vertical direction and $N_{\rm{h}}$ antennas in the horizontal direction. The number of transmit antennas is $N_{\rm{t}}=N_{\rm{v}}N_{\rm{h}}$. Suppose that the BS is serving $K$ users with single antenna simultaneously, and the CSI in spatial domain for user $k$ is ${\bf{H}}_k^{\rm{s}} \in \mathbb{C}^{{N_{\rm{v}}} {\times} {N_{\rm{h}}}}$.

The 3D-MIMO channel exhibits sparse property when transformed to angular domain. Denote the vector form of spatial domain channel as ${{\bf{h}}_k^{\rm{s}}} = vec({\bf{H}}_k^{\rm{s}})$ for user $k$. ${{\bf{h}}_k^{\rm{s}}}$ can be equivalently transformed to angular domain ${{\bf{h}}_k^{\rm{a}}}={{\bf{h}}_k^{\rm{s}}} \bf{U}$, where ${\bf{U}} \in \mathbb{C}^{{N_{\rm{t}}} {\times} {N_{\rm{t}}}}$ has $N_{\rm{t}}$ mutually orthogonal beams which constitute the basis of the angular domain. These orthogonal basis provide a spatial decomposition of the total transmit signal into the multi-beams along the different physical directions. Therefore, BS with large antenna array can provide high spatial resolution in 3D-MIMO systems \cite{fscsb}. Since the number of channel paths is limited \cite{ccsds}, the channel is approximately sparse in angle domain on sufficient spatial resolution. In the rest of the paper, the superscript $\rm{a}$ is omitted for concision.

The BS transmits signals to $k$ users simultaneously. The received signal at user $k$ is given by

\begin{equation}\label{reck}
{y_k} = {\bf{h}}_k^H{{\bf{w}}_k}{x_k} + \sum\limits_{j = 1,j \ne k}^K {{\bf{h}}_k^H{{\bf{w}}_j}{x_j}}  + {n_k},
\end{equation}

\noindent where ${{\mathbf{w}}_{k}}\in {{\mathbb{C}}^{N_{\rm{t}}\times 1}}$ is the beamforming vector of user $k$. Denote ${\bf{W}} = [{\bf{w}}_1,{\bf{w}}_2...,{\bf{w}}_K]$ as the beamforming matrix. ${{x}_{j}}$ is the transmit symbol for user $k$, which satisfies ${\rm E}[x_j^*{x_j}] = 1$, ${{n}_{k}}$ is the zero-mean complex Gaussian white noise with the variance of ${{\sigma }_{n}^2}$. SINR of user $k$ can be expressed as

\begin{equation}\label{sinrk}
{\rm{SINR}}_k = \dfrac{{{{\left| {{\bf{h}}_k^H{{\bf{w}}_k}} \right|}^2}}}{{\sum\limits_{j = 1,j \ne k}^K {{{\left| {{\bf{h}}_k^H{{\bf{w}}_j}} \right|}^2}}  + {\sigma}_n^2}}.
\end{equation}

Practically, the BS only knows the imperfect CSI. Denote the estimated sparse channel for user $k$ as ${{\hat{\bf{h}}}_k}$, and define the CSI error for user $k$ as

\begin{equation}\label{delta}
 {\bm{\delta}}_k={\bf{h}}_k - {{\hat{\bf{h}}}_k}.
\end{equation}

The commonly used set for channel error ${\bm{\delta}}_k$ is assumed to be covered by a spherical region. Specifically, the CSI uncertainty set is

\begin{equation}\label{unc-l2}
 \mathcal{H}_k^2=\{ {{\hat{\bf{h}}}_k}+{\bm{\delta}}_k \mid {\left\| {\bm{\delta}}_k \right\|}_2 \le {\epsilon}_k \}.
\end{equation}

\noindent where ${\epsilon}_k$ is the bound of the uncertainty. However in 3D-MIMO systems, both ${\bf{h}}_k$ and ${{\hat{\bf{h}}}_k}$ are sparse. Suppose that the sparsity of ${\bf{h}}_k$ and ${{\hat{\bf{h}}}_k}$ are $s$, i.e. there are at most $s$ non-zero elements in ${\bf{h}}_k$ and ${{\hat{\bf{h}}}_k}$, then the sparsity of CSI error ${\bm{\delta}}_k$ is $2s$, which is still sparse in 3D-MIMO systems. Thus the estimation errors should be bounded by $l_0$-norm, i.e. ${\left\| {\bm{\delta}}_k \right\|}_0 \le {\epsilon}_k$. However, $l_0$-norm based constraint is non-convex and intractable. In general, $l_0$-norm  is often replaced by $l_1$-norm, since $l_1$-norm based constraint is not only convex but can capture the sparse feature as well \cite{cs}. Thus, we define the $l_1$-norm bounded CSI uncertainty as

\begin{equation}\label{unc-l1}
 \mathcal{H}_k^1=\{ {{\hat{\bf{h}}}_k}+{\bm{\delta}}_k \mid {\left\| {\bm{\delta}}_k \right\|}_1 \le {\epsilon}_k \}.
\end{equation}

Given the channel uncertainty in (\ref{unc-l1}), the robust beamforming problem is to design a beamforming matrix which minimizes the transmit power required to ensure that the users' worst case SINR constraints are satisfied. The problem can be formulated as
\begin{align}\label{P0}
\mathop {\min }\limits_{{{\bf{w}}_k}} \;\; &  \sum\limits_{k = 1}^K {{{\left\| {{{\bf{w}}_k}} \right\|}_2^2}} \\
{\rm{ s}}{\rm{.t}}{\rm{.}}\;\; &  {\rm{SINR}}_k \ge {{\gamma}_k},\;\; \forall {\bf{h}}_k \in \mathcal{H}_k^1, \;\;k = 1,2,...,K, \nonumber
\end{align}

\noindent where ${{\gamma }_{k}}$ is the predefined worst case SINR targets that each user shall achieve.

\section{Algorithm}

Note that when the uncertainty set $\mathcal{H}_k^1$ is replaced with the set $\mathcal{H}_k^2$, the problem (\ref{P0}) can be recasted as the robust beamforming problem with spherical uncertainty, as proposed in \cite{sph1}. It can be solved by semi-definite programming (SDP) with the aid of \emph{S-procedure} \cite{convexop}. The following theorem shows that the minimum transmit power required for the robust beamforming problem (\ref{P0}) with $l_1$-norm bounded uncertainty outperforms that with spherical bounded uncertainty.

\emph{Theorem 1:} For any given ${\gamma}_k$ and ${\epsilon}_k$, the optimal objective value of problem (\ref{P0}) with uncertainty set $\mathcal{H}_k^1$ is smaller than that with uncertainty set $\mathcal{H}_k^2$.

\emph{Proof:} The worst case SINR constraint in (\ref{P0}) can be equivalently written as

\begin{equation}
\mathop {\min}\limits_{{\bf{h}}_k \in \mathcal{H}_k^i} \; \dfrac{{{{\left| {{\bf{h}}_k^H{{\bf{w}}_k}} \right|}^2}}}{{\sum\limits_{j = 1,j \ne k}^K {{{\left| {{\bf{h}}_k^H{{\bf{w}}_j}} \right|}^2}}  + {\sigma}_n^2}} \ge {{\gamma}_k},\;\forall k,
\label{P0con}
\end{equation}

\noindent where $i=1,2$ denote two uncertainty sets. (\ref{P0con}) constraints the feasible set of $\left\{ {\bf{w}}_k \right\}_{k=1}^{K}$, such that the minimum SINR exceeds the target ${{\gamma}_k}$ for all channels ${\bf{h}}_k$ within the uncertainty set $\mathcal{H}_k^i$. It's easy to see that the larger the uncertainty region is, the smaller the feasible set is.

On the other hand, it is obvious that for any given vector $\bf{x}$, ${\left\| {\bf{x}} \right\|}_1 \ge {\left\| {\bf{x}} \right\|}_2$. Thus, for a given bound ${\epsilon}_k$, we have

\begin{equation}
\mathcal{H}_k^1 \subseteq \mathcal{H}_k^2,
\label{H1H2}
\end{equation}

\noindent which means the $l_1$-norm bounded uncertainty set is contained in the spherical bounded uncertainty set. From (\ref{P0con}) and (\ref{H1H2}), we have

\begin{equation}
\mathop {\min}\limits_{{\bf{h}}_k \in \mathcal{H}_k^1} \; {\rm{SINR}}_k \ge \mathop {\min}\limits_{{\bf{h}}_k \in \mathcal{H}_k^2} \; {\rm{SINR}}_k \ge {{\gamma}_k}, \;\forall k,
\label{contain}
\end{equation}

\noindent which means for any given ${{\gamma}_k}$, the feasible set of (\ref{P0con}) with uncertainty set $\mathcal{H}_k^2$ is contained in the feasible set with $\mathcal{H}_k^1$. Thus the optimal objective value of problem (\ref{P0}) with the uncertainty set $\mathcal{H}_k^1$ is smaller than that with the uncertainty set $\mathcal{H}_k^2$.\hfill $\blacksquare$

The problem (\ref{P0}) is intractable due to that the channel uncertainty contains infinite constraints. By introducing an auxiliary variable $p$, (\ref{P0}) can be equivalently reformulated as
\begin{subequations}
\begin{align}\label{P1}
\mathop {\min }\limits_{{{\bf{w}}_k},p} \;\; & p \\
{\rm{ s}}{\rm{.t}}{\rm{.}} \;\; & \sum\limits_{k = 1}^K {{{\left\| {{{\bf{w}}_k}} \right\|}_2^2}} \le p \label{P1-b}\\
& {\rm{SINR}}_k \ge {{\gamma}_k}\;, \label{P1-c} \\
& \;\;\;\; \forall \; {\bf{h}}_k \in \mathcal{H}_k^1\;, \;\forall k, \nonumber
\end{align}
\end{subequations}

\noindent where (\ref{P1-b}) can be converted to an SOC constraint. From the definition of SINR (\ref{sinrk}), the constraint ${\rm{SINR}}_k \ge {{\gamma}_k}$ in (\ref{P1-c}) can be rewritten as

\begin{equation}\label{sinrequ}
{\left\| {[{{\bf{h}}_k^H} {\bf{W}},\;{\sigma_n}]} \right\|}_2 \le {\beta _k} \left| {{{\bf{h}}_k^H} {{\bf{w}}_k}} \right|,
\end{equation}

\noindent where ${{\beta}_k}=\sqrt{1+1/{{\gamma}_k}}$. Thus, (\ref{P1-c}) can be relaxed as

\begin{subequations}
\begin{align}
& \mathop {\min}\limits_{{\bf{h}}_k \in \mathcal{H}_k^1} \;  \left| {{{\bf{h}}_k^H} {{\bf{w}}_k}} \right| \ge \dfrac{t_k}{{\beta}_k},  \label{P2-a}\\
& \mathop {\max}\limits_{{\bf{h}}_k \in \mathcal{H}_k^1} \; {\left\| {[{{\bf{h}}_k^H} {\bf{W}},\;{\sigma_n}]} \right\|}_2 \le {t_k}, \;\forall k, \label{P2-b}
\end{align}
\end{subequations}

\noindent where $t_k$ is a new optimization variable. Since (\ref{P2-a}) and (\ref{P2-b}) share the same CSI uncertainty, the worst case constraints (\ref{P1-c}) cannot be decoupled into finding the minimum numerator and the maximum denominator independently. Thus, (\ref{P2-a}) and (\ref{P2-b}) are relaxed constraints from (\ref{P1-c}), and can be considered as a lower bound for constraints (\ref{P1-c}).

To deal with the constraint (\ref{P2-a}), we need to find the minimum value of $\left| {{{\bf{h}}_k^H} {{\bf{w}}_k}} \right|$ for all the ${\bf{h}}_k \in \mathcal{H}_k^1$. Consider the left-hand side of (\ref{P2-a})

\begin{align}\label{atmp1}
\left| {{{\bf{h}}_k^H} {{\bf{w}}_k}} \right| & =   \left|  ({{\hat{\bf{h}}}_k^H}+{\bm{\delta}_k^H}) {{\bf{w}}_k} \right|\\
& \ge \left| {{{\hat{\bf{h}}}_k^H} {{\bf{w}}_k}} \right|-\left| {{\bm{\delta}_k^H} {{\bf{w}}_k}} \right| \nonumber
\end{align}

\noindent where the inequality is due to the triangle inequality of vector norm. Denote the $n$-th element of a vector $\bf{x}$ as $x(n)$, then

\begin{align}\label{atmp2}
\left| {{\bm{\delta}_k^H} {{\bf{w}}_k}} \right| & = \left| \sum\limits_{n=1}^{N_{\rm{t}}} {{\delta}_k^*}(n) w_k(n) \right|\\
& \le \sum\limits_{n=1}^{N_{\rm{t}}} \left| {{\delta}_k^*}(n) w_k(n) \right| \nonumber \\
& \le {{\eta}_k} \sum\limits_{n=1}^{N_{\rm{t}}} \left| {\delta_k}(n) \right| \le {\epsilon}_k {{\eta}_k}, \nonumber
\end{align}

\noindent where ${{\eta}_k} = {\left\| {\bf{w}}_k \right\|}_{\infty}$ is the infinity norm of ${\bf{w}}_k$. Since $\left| {{{\hat{\bf{h}}}_k^H} {{\bf{w}}_k}} \right| \ge \text{Re} ({{{\hat{\bf{h}}}_k^H} {{\bf{w}}_k}}) $, from (\ref{atmp1}) and (\ref{atmp2}), we can rewrite (\ref{P2-a}) as the set of following constraints

\begin{equation} \label{aequ}
\left\{\begin{array}{l}
\text{Re} ({{{\hat{\bf{h}}}_k^H} {{\bf{w}}_k}}) - {\epsilon}_k \eta  \ge \dfrac{t_k}{\beta}\\
{\left\| {\bf{w}}_k \right\|}_{\infty} \le \eta \;\;\;\;  k=1,2,...,K.
\end{array}\right.,
\end{equation}

\noindent where $\eta = \max {\left\{ {\eta}_k \right\}}_{k=1}^K$, is a new optimization variable. Note that (\ref{aequ}) are linear constraints.

Similarly, we have to find the maximum value of ${\left\| {{\bf{h}}_k^H} {\bf{W}} \right\|}_2$ for all the ${\bf{h}}_k \in \mathcal{H}_k^1$. The left-hand side of (\ref{P2-b}) is

\begin{equation}\label{btmp1}
{\left\| {{\bf{h}}_k^H} {\bf{W}} \right\|}_2  \le {\left\| {{\hat{\bf{h}}}_k^H} {\bf{W}} \right\|}_2 + {\left\| {{\bm{\delta}}_k^H} {\bf{W}} \right\|}_2.
\end{equation}

Suppose that ${\bf{v}}(n) \in \mathbb{C}^{1 \times K}$ is the $n$-th row of matrix ${\bf{W}}$. Assuming that ${\left\| {\bf{v}}(n) \right\|}_2 \le \alpha$ for any $1 \le n \le N_{\rm{t}}$, then

\begin{align}\label{btmp2}
{\left\| {{\bm{\delta}}_k^H} {\bf{W}} \right\|}_2 & = {\left\| \sum\limits_{n=1}^{N_{\rm{t}}} {\delta_k^*}(n) {\bf{v}}(n)  \right\|}_2\\
& \le \sum\limits_{n=1}^{N_{\rm{t}}} {\left\| {\delta_k^*}(n) {\bf{v}}(n) \right\|}_2  \le \sum\limits_{n=1}^{N_{\rm{t}}} {\left| {\delta_k^*}(n) \right|} {\left\| {\bf{v}}(n) \right\|}_2  \nonumber \\
& \le \alpha \sum\limits_{n=1}^{N_{\rm{t}}} \left| {\delta_k}(n) \right| \le {\epsilon}_k \alpha. \nonumber
\end{align}

From (\ref{btmp1}) and (\ref{btmp2}), we can rewrite (\ref{P2-b}) as

\begin{equation} \label{bequ}
\left\{\begin{array}{l}
{\left\| [{{\hat{\bf{h}}}_k^H} {\bf{W}},\;{\alpha}{\epsilon}_k,\;{\sigma}_n] \right\|}_2 \le t_k,\;  \forall k, \\
{\left\| {\bf{v}}(n) \right\|}_2 \le \alpha,\;  \forall n.
\end{array}\right.,
\end{equation}

\noindent which are SOC constraints. Finally, we arrive at the relaxed reformulation of (\ref{P0})

\begin{align}\label{final}
\mathop {\min }\limits_{{{\bf{w}}_k},p,t_k,\eta,\alpha} \;\; &  p \\
{\rm{ s}}{\rm{.t}}{\rm{.}}\;\; & \sum\limits_{k = 1}^K {{{\left\| {{{\bf{w}}_k}} \right\|}_2^2}} \le p \nonumber \\
& (\ref{aequ}), (\ref{bequ}). \nonumber
\end{align}

Note that all the constraints in (\ref{final}) are linear or SOCs. Thus it can be solve by convex optimization toolboxes, such as the CVX toolbox \cite{cvx}.

\begin{table}
  \centering
  \caption{Physical Channel Model Parameters}\label{tab1}
  \begin{tabular}{c|c}
    \hline
    \textbf{Parameters} & \textbf{Configurations} \\
    \hline
    Num of channel taps & $L=6$ \\
    Antenna spacing & 0.5$\lambda$\\
    Vertical Angle of Departure & $\theta \sim \mathcal{U}(0,{\pi}/2)$ \\
    Horizontal Angle of Departure & $\phi \sim \mathcal{U}(0,{\pi})$ \\
    Path loss per channel tap & $\beta \sim \mathcal{U}(0,1)$ \\
    \hline
  \end{tabular}
\end{table}

\section{Simulation Results}
We consider a downlink multi-user 3D-MIMO system where the BS is equipped with $N_{\rm{v}} \times N_{\rm{h}}=4 \times 8$ antennas, and serves $K=4$ single antenna active users simultaneously. The sparse channels are generated through physical channel model \cite{tse2005}. The detailed parameters are listed in Table \ref{tab1}, where $\lambda$ is carrier wavelength, and $x \sim \mathcal{U}(a,b)$ means parameter $x$ is randomly generated from uniform distribution within the interval $[a,b]$. The CSI uncertainty bound is ${\epsilon}_k={\epsilon}{\left\| {\bf{h}} \right\|}_2$ for a given $\epsilon$, and the noise power is ${\sigma}_n^2=0.1{\left\| {\bf{h}} \right\|}_2^2$.

We compare the proposed robust beamforming scheme with two other strategies. Perfect CSI shows the performance when the perfect CSI is used to design the optimal beamforming matrix \cite{bengtsson2001optimal}. Robust SDP is the algorithm proposed in \cite{sph1}, with the spherical uncertainty set (\ref{unc-l2}). The simulations are averaged over $N_{\rm{run}}=100$ runs, each of which is chosen so that the problem is feasible for the observed range of the constraints.

\begin{figure}[htb]
    {
    \setlength{\abovecaptionskip}{0pt} 
    \setlength{\belowcaptionskip}{10pt} 
    \centering{
    \includegraphics[width=3 in]{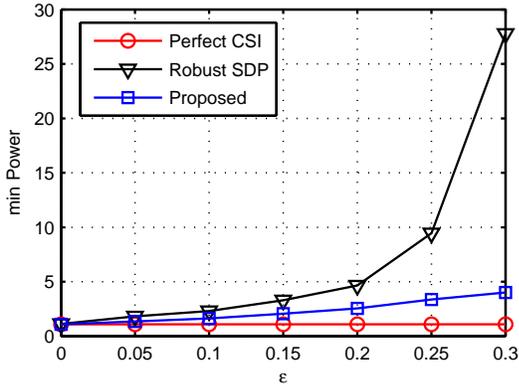}
    \caption{Minimum transmit power versus uncertainty bound ${\epsilon}$, ${{\gamma}_k}=3\rm{dB}$.}}
    \label{Figure 1}
    }
\end{figure}

\begin{figure}[htb]
    {
    \setlength{\abovecaptionskip}{0pt} 
    \setlength{\belowcaptionskip}{10pt} 
    \centering{
    \includegraphics[width=3 in]{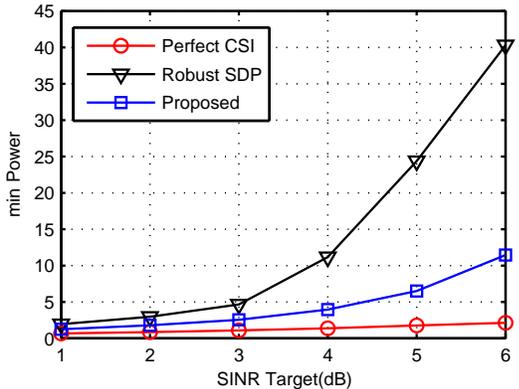}
    \caption{Minimum transmit power versus SINR target, $\epsilon=0.2$.}}
    \label{Figure 2}
    }
\end{figure}

Fig.1 plots the minimum transmit power against the uncertainty bound $\epsilon$. The SINR target is ${{\gamma}_k}=3\rm{dB}$ for all users. It can be seen that when $\epsilon=0$, which means the perfect CSI is available at BS, all three algorithms have the same performance. When the uncertainty bound goes larger, the required minimum power of Robust SDP algorithm grows much faster than the proposed strategy. That is because the spherical uncertainty region spans faster than $l_1$-norm bounded uncertainty region as $\epsilon$ increases.

Fig.2 plots the minimum transmit power against the SINR targets, and the uncertainty bound is $\epsilon=0.2$. It can be seen that the required minimum power of Robust SDP algorithm scales significantly when the SINR targets grows larger, while the required minimum power of the perfect CSI case and the proposed strategy grows much slower.

\section{Conclusion}
In this paper, we investigate a novel robust beamforming scheme in 3D-MIMO systems, where a $l_1$-norm bounded channel uncertainty model is proposed to describe the sparse channel errors. The problem is reformulated as an SOCP, and is proved to outperform the method with conventional spherical uncertainty. Simulation results verify that the proposed scheme achieves better performance than conventional spherical uncertainty based scheme with respect to various SINR targets and uncertainty bounds. The idea of this paper is not only suitable for the robust beamforming problems in 3D-MIMO systems, but also can be applied to robust beamforming problems in other massive MIMO systems with sparse perturbation.

\section*{ACKNOWLEDGMENT}
This work was supported by the NSF of China (Grant No. 61501124, No. 71731004), and the National Key Research and Development Program of China (No.213).





\bibliographystyle{IEEEtran}
\bibliography{Myreference}

\begin{thebibliography}{10}
\providecommand{\url}[1]{#1}
\csname url@samestyle\endcsname
\providecommand{\newblock}{\relax}
\providecommand{\bibinfo}[2]{#2}
\providecommand{\BIBentrySTDinterwordspacing}{\spaceskip=0pt\relax}
\providecommand{\BIBentryALTinterwordstretchfactor}{4}
\providecommand{\BIBentryALTinterwordspacing}{\spaceskip=\fontdimen2\font plus
\BIBentryALTinterwordstretchfactor\fontdimen3\font minus
  \fontdimen4\font\relax}
\providecommand{\BIBforeignlanguage}[2]{{%
\expandafter\ifx\csname l@#1\endcsname\relax
\typeout{** WARNING: IEEEtran.bst: No hyphenation pattern has been}%
\typeout{** loaded for the language `#1'. Using the pattern for}%
\typeout{** the default language instead.}%
\else
\language=\csname l@#1\endcsname
\fi
#2}}
\providecommand{\BIBdecl}{\relax}
\BIBdecl

\bibitem{fdmimo}
H.~Ji, Y.~Kim, J.~Lee, E.~Onggosanusi, Y.~Nam, J.~Zhang, B.~Lee, and B.~Shim,
  ``Overview of full-dimension mimo in lte-advanced pro,'' \emph{IEEE Commun.
  Mag.}, vol.~55, no.~2, pp. 176--184, February 2017.

\bibitem{ncwun}
T.~Marzetta, ``Noncooperative cellular wireless with unlimited numbers of base
  station antennas,'' \emph{IEEE Trans. Wireless Commun}, vol.~9, no.~11, pp.
  3590--3600, November 2010.

\bibitem{com3d}
X.~Cheng, B.~Yu, L.~Yang, J.~Zhang, G.~Liu, Y.~Wu, and L.~Wan, ``Communicating
  in the real world: 3d mimo,'' \emph{IEEE Wireless Commun.}, vol.~21, no.~4,
  pp. 136--144, 2014.

\bibitem{sph1}
F.~Wang, Y.~Huang, W.~X, and Y.~Zhu, ``Robust beamforming designs for multiuser
  miso downlink with per-antenna power constraints,'' \emph{EURASIP Journal on
  Wireless Commun. and Netw.}, vol. 2015, no.~1, p. 204, 2015.

\bibitem{sph3}
J.~Liao, M.~R.~A. Khandaker, and K.~K. Wong, ``Robust power-splitting swipt
  beamforming for broadcast channels,'' \emph{IEEE Commun. Lett.}, vol.~20,
  no.~1, pp. 181--184, Jan 2016.

\bibitem{elli1quan1}
A.~Pascual-Iserte, D.~P. Palomar, A.~I. Perez-Neira, and M.~A. Lagunas, ``A
  robust maximin approach for mimo communications with imperfect channel state
  information based on convex optimization,'' \emph{IEEE Trans. Signal
  Process.}, vol.~54, no.~1, pp. 346--360, 2006.

\bibitem{sph2quan2}
N.~Vucic and H.~Boche, ``Robust qos-constrained optimization of downlink
  multiuser miso systems,'' \emph{IEEE Trans. Signal Process.}, vol.~57, no.~2,
  pp. 714--725, 2009.

\bibitem{inteelli2}
M.~F. Hanif, L.~N. Tran, A.~Tolli, M.~Juntti, and S.~Glisic, ``Efficient
  solutions for weighted sum rate maximization in multicellular networks with
  channel uncertainties,'' \emph{IEEE Trans. Signal Process.}, vol.~61, no.~22,
  pp. 5659--5674, 2013.

\bibitem{severalunc}
J.~Wang, M.~Bengtsson, B.~Ottersten, and D.~P. Palomar, ``Robust mimo precoding
  for several classes of channel uncertainty,'' \emph{IEEE Trans. Signal
  Process.}, vol.~61, no.~12, pp. 3056--3070, 2013.

\bibitem{ccsds}
W.~Bajwa, J.~Haupt, A.~Sayeed, and R.~Nowak, ``Compressed channel sensing: A
  new approach to estimating sparse multipath channels,'' \emph{Proc. IEEE},
  vol.~98, no.~6, pp. 1058--1076, 2010.

\bibitem{sce1}
M.~Masood, L.~H. Afify, and T.~Y. Al-Naffouri, ``Efficient coordinated recovery
  of sparse channels in massive mimo,'' \emph{IEEE Trans. Signal Process.},
  vol.~63, no.~1, pp. 104--118, 2015.

\bibitem{ssce}
K.~Liu, H.~Feng, T.~Yang, and B.~Hu, ``Structured sparse channel estimation for
  3d-mimo systems,'' in \emph{Proc. IEEE Veh. Technol. Conf.}, 2016, pp. 1--6.

\bibitem{fscsb}
H.~Feng, S.~Zhang, K.~Liu, T.~Yang, and B.~Hu, ``From sparse channel to sparse
  beamforming: A 3d-mimo case,'' in \emph{Proc. IEEE Global Commun. Conf.},
  2015.

\bibitem{cs}
D.~L. Donoho, ``Compressed sensing,'' \emph{IEEE Trans. Inf. Theory}, vol.~52,
  no.~4, pp. 1289--1306, 2006.

\bibitem{convexop}
S.~Boyd and L.~Vandenberghe, \emph{Convex Optimization}.\hskip 1em plus 0.5em
  minus 0.4em\relax New York, NY, USA: Cambridge University Press, 2004.

\bibitem{cvx}
M.~Grant and S.~Boyd, ``{CVX}: Matlab software for disciplined convex
  programming, version 2.1,'' Available: \url{http://cvxr.com/cvx}, Mar. 2014.

\bibitem{tse2005}
D.~Tse and P.~Viswanath, \emph{Fundamentals of wireless communication}.\hskip
  1em plus 0.5em minus 0.4em\relax Cambridge university press, 2005.

\bibitem{bengtsson2001optimal}
M.~Bengtsson and B.~Ottersten, ``Optimal downlink beamforming using
  semidefinite optimization,'' in \emph{Proc. 37th annu. allerton}, 1999, pp.
  987--996.

\end{thebibliography}
%

%
%

\end{document}